\documentclass[12pt,dvips]{article}
\usepackage{epsfig}

\newcommand{\be}{\begin{equation}}
\newcommand{\ee}{\end{equation}}
\newcommand{\condensate}{\ensuremath{\langle\bar{\psi}\psi\rangle}{}}

\newcommand{\psibar}{\bar{\psi}}
\newcommand{\ubar}{\bar{u}}
\newcommand{\dbar}{\bar{d}}
\newcommand\DWI{\ensuremath{{\cal D}_\textrm{\scriptsize Wi}\,}{}}
\newcommand\DFP{\ensuremath{{\cal D}_\textrm{\scriptsize Fp}\,}{}}
\newcommand\DOV{\ensuremath{{\cal D}_\textrm{\scriptsize Ne}\,}{}}
\newcommand\KNE{\ensuremath{\kappa_\textrm{\scriptsize Ne}}{}}
\newcommand\unitmatrix{\ensuremath{\textrm{\boldmath{$\mathsf{1}$}}}}

\begin{document}

\thispagestyle{empty}

\date{\today}
\title{
\vspace{-5.0cm}
\begin{flushright}
{\normalsize UNIGRAZ-}\\
{\normalsize UTP-03-09-98}\\
\end{flushright}
\vspace*{2.5cm}
Wilson, fixed point
and Neuberger's lattice Dirac operator for the \\ Schwinger model
\thanks{Supported by Fonds 
zur F\"orderung der Wissenschaftlichen Forschung 
in \"Osterreich, Project P11502-PHY.}  }
\author
{\bf F. Farchioni, 
I. Hip and C. B. Lang \\  \\
Institut f\"ur Theoretische Physik,\\
Universit\"at Graz, A-8010 Graz, AUSTRIA}
\maketitle
\begin{abstract}
We perform a comparison between different lattice regularizations
of the Dirac operator for massless fermions in the framework
of the single and two flavor Schwinger model. We consider 
a) the Wilson-Dirac operator at the critical value of the
hopping parameter;  b)  Neuberger's 
overlap operator; c) the fixed point operator.
We test chiral properties of the spectrum, dispersion relations and
rotational invariance of the mesonic bound state propagators.

\end{abstract}



\newpage

\section{Motivation and Introduction}

The Nielsen-Ninomiya \cite{NiNi81} theorem and the Ginsparg-Wilson
\cite{GiWi82} condition (GWC) provide us with the crucial information
under which circumstances \cite{Ha98c} remnants of chiral symmetry 
may stay with a lattice action for fermions,
like it is e.g. the case for overlap fermions \cite{NaNe939495} or 
fixed point actions \cite{HaLaNi98}. Recently L\"uscher
\cite{Lu98} has pointed out the explicit form of the
underlying symmetry and indicated possible generalizations.

Fermion actions for massless fermions satisfying the GWC
\be
\frac{1}{2}\left\{\, {\cal D},\gamma^5\,\right\}=
a\,\left(\,{\cal D}\,\gamma^5\,R\,{\cal D}\,\right)\;\;,
\label{eq:gwcg}
\ee
where  ${\cal D}$ is the lattice Dirac operator, violate 
chiral symmetry up to a local term ${\cal O}(a)$;
the r.h.s. of the above equation is local since $R$ is.

It was pointed out in \cite{Ha98c} that actions which are 
fixed points under real space renormalization group (block spin) 
transformations (BST), are solutions of the GWC.
$R$ is then local and bounded and as a consequence 
the spectrum of ${\cal D}$ in complex space is confined between two
circles \cite{HaLaNi98}:
\be\label{CirclesBound}
|\lambda-r_{min}|\geq r_{min} \;,\quad |\lambda-r_{max}|\leq r_{max}\;\;,
\ee
where the real numbers $r_{min}$ and $r_{max}$ are related to the maximum and minimum 
eigenvalue of $R$ respectively.
For non-overlapping BSTs $R=\frac{1}{2}$ and (\ref{CirclesBound})
reduces to $|\lambda-1|=1$, i.e. the spectrum lies on circle.

Independent solutions of the GWC are provided by the overlap 
formalism \cite{NaNe939495}, which allows the formulation 
of chiral fermions on the lattice. These solutions are obtained in
an elegant way, as shown recently by Neuberger \cite{Ne98,Ne98a}, 
through some projection of the Wilson operator with negative 
fermion mass. Even in this case we have $R=\frac{1}{2}$ and
$|\lambda-1|=1$.

For the Schwinger model (2D QED) we are in a situation where we have
access to three different lattice Dirac operators for massless fermions,
namely the original Wilson operator \DWI at $\kappa_c(\beta)$,
the Neuberger-projected operator \DOV and a numerically determined
and therefore approximate fixed point operator 
\DFP \cite{LaPa98b,FaLaWo98}. Studying these alternatives we may ask the
following questions:
\begin{itemize}
\item For given gauge field configurations:
What is the relation between real eigenvalues of the lattice Dirac operator
and the geometric
topological charge? To what extent is the Atiyah-Singer Index Theorem
(ASIT) realized in these lattice environments?
\item In the continuum, the eigenvalue distribution is related 
to the condensate $\condensate$ (Banks-Casher formula \cite{BaCa80}).
What about the lattice theory for the given Dirac operators? 
\item The first two questions concern chiral symmetry and the 
phenomenon of fermion condensation.
Important for the eventual study of the continuum limit of the full
theory are also spectral properties: What is the
behavior of e.g. dispersion relations?
Concerning off-shell properties, what about recovery of rotational invariance
of the propagators?
\end{itemize}

Here we perform a Monte Carlo simulation for the one- and two-flavor
Schwinger model with gauge fields in the compact representation and the
three different lattice Dirac operators. For the two-flavor model one
expects a massless bound state in the chiral limit, which is of
particular interest in this framework.  We stress that the ensemble of
gauge configurations used is the same in all three cases, the sampling
being performed using the one-plaquette standard gauge action.  The
unquenching is obtained through the multiplication by the fermion
determinant.

\section{Lattice Dirac Operators}

The three fermion actions may be written $\psibar\,{\cal D}\,\psi$,
where the fields at each site are two-components Grassmann
variables $\psibar, \psi$ and  ${\cal D}$ is a matrix 
in Euclidean and Dirac space (lattice Dirac operator).
For two flavors the number of fields duplicate and the action
becomes $\ubar\,{\cal D}\,u+\dbar\,{\cal D}\,d$ (with independent
Grassmann fields $\ubar$, $u$, $\dbar$, $d$). 

All three Dirac operators are non-hermitian but
have $\gamma_5$-hermiticity:\goodbreak 
\noindent$\gamma_5{\cal D}\gamma_5={\cal D}^{\dagger}$. 
Their eigenvalue spectrum is therefore symmetric
with regard to complex conjugation. 

\subsection{Wilson Dirac operator}
We write the Wilson Dirac operator in the form
\be\label{DWI}
\DWI(x,y)=(m+D)\,\unitmatrix_{xy} - \frac{1}{2} \sum_{\mu} \left[
 (1+\sigma_\mu) \,U_{xy}\,\delta_{x,y-\mu}
+ (1-\sigma_\mu)\, U_{yx}^\dagger\,\delta_{x,y+\mu}\right]\;.
\ee
The hopping parameter $\kappa$ is related to the (bare) quark mass 
$m$ through the relation $\kappa=1/(2m+2D)$.
 
The chiral limit is obtained in this environment 
for $\kappa\rightarrow\kappa_c(\beta)$, where 
$\kappa_c(\beta)$ should be determined in some way (see the following).
Thus we will work at $D=2$, and $\kappa=\kappa_c(\beta)$
(corresponding massless quarks) as discussed below.

\subsection{Neuberger's operator}
\label{sec:Neuberger}

Neuberger suggests \cite{Ne98} to start with the Wilson Dirac operator
at some value of $m \in (-1,0)$ corresponding to
$\frac{1}{2D}<\kappa<\frac{1}{2D-2}$
and then construct 
\be\label{DOV}
\DOV = \unitmatrix  + \gamma_5\, \epsilon(\gamma_5\,\DWI) \;.
\ee
We call the actual value of $\kappa$ used in the above definition
$\KNE$.  Some words about the choice of $\KNE$: according to
\cite{Ne98a} it is arbitrary, in the sense that any (strictly negative)
value of $m$ in the interval $(-1,0)$ reproduces the correct continuum
theory (see also the discussion on a suitable choice of $\KNE$ in
\cite{Ch98a}), but it may be optimized with regard to its scale
dependence by looking for example at the behavior of the (projected)
spectrum.  Comparing expectation values of operators like $\condensate$
for different $\KNE$ one has to take care of the proper
normalization \cite{KiNaNe98}.  One may see from the comparison with
free lattice fermions that there is a (trivial) factor of $\sqrt{m}$,
i.e.  $\condensate=\condensate_\textrm{\scriptsize Ne}/|m|$ in our
convention.\footnote{We thank H. Neuberger for pointing this out to
us.} Whenever not mentioned otherwise we choose
$\KNE=\frac{1}{2}$ ($m=-1$) for our exploration; we also discuss
results with some smaller values.

The operative definition of  $\epsilon(\gamma_5\,\DWI)$ entering
the above equation is:
\be\label{defeps}
\epsilon(\gamma_5\,\DWI)= 
U \, \textrm{Sign}(\Lambda)\, U^\dagger \quad \textrm{with}\; 
\gamma_5\,\DWI = U \, \Lambda \,U^\dagger\;.
\ee
Here $\textrm{Sign}(\Lambda)$ denotes the diagonal matrix containing 
the signs of the eigenvalue matrix $\Lambda$
obtained through the unitary transformation $U$ of the hermitian 
matrix $\gamma_5\,\DWI$.
There are various efficient ways to numerically find \DOV
without passing through  the diagonalization problem,  
prohibitive for $D=4$ 
\cite{NeEdHeNa98} (for $D=2$, see also \cite{Ch98}).
In our simple context computer time is no real obstacle
and therefore we use the direct definition (\ref{defeps}),
explicitly performing the diagonalization.

As observed in \cite{Ne98a}, gauge configurations with non-zero
topological charge imply  exact zero eigenvalues of $\DOV$. The
subsequent inversion, necessary to find the quark propagators, is then
not possible.  The correct way to proceed is to introduce a regulator
cutoff-mass:  ${\cal D}\rightarrow{\cal D}+\mu\,\unitmatrix$ and then consider
the limit $\mu\rightarrow0$.

\subsection{Fixed point Dirac operator}

In \cite{LaPa98b} the fixed point Dirac operator was parameterized as
\be\label{DFP}
\DFP(x,y) =
\sum_{i=0}^3\sum_{x\, , f}\,  \rho_i(f)\, 
\sigma_i\, U(x,f)\;,\quad\textrm{with} \; y\equiv x+\delta f\;.
\ee
Here $f$ denotes a closed
loop through $x$ or a path from the lattice site $x$ to $y=x+\delta f$
(distance vector $\delta f$) and $U(x,f)$ is the parallel transporter
along this path. The $\sigma_i$-matrices denote the Pauli matrices for
$i= 1,2,3$ and the unit matrix for $i=0$. 
The action obeys the usual symmetries as discussed in \cite{LaPa98b};
altogether it has 429 terms per site.
The action was determined for gauge fields distributed according to the
non-compact formulation with the Gaussian measure. There excellent 
scaling properties, rotational invariance and continuum-like dispersion
relations were observed at various values of the gauge coupling $\beta$.

In \cite{FaLaWo98} the action was studied both, for compact and the
original non-compact gauge field distributions.  In the compact case
the action is not expected to exactly reproduce the fixed point of the
corresponding BST, but nevertheless it is still a solution of the GWC;
violations of the GWC are instead introduced by the parameterization
procedure, which cuts off the less local couplings.  We demonstrated
that indeed the spectrum is close to circular, somewhat fuzzy at small
values of $\beta\leq 2$ but excellently living up to the theoretical
expectations of \cite{HaLaNi98} at large gauge couplings $\beta\geq
4$.

Here we study the action only for the compact gauge field distributions
in order to allow a direct comparison with the other lattice Dirac
operators.

\section{Simulation Details}

Uncorrelated gauge configurations have been generated in the quenched
setup.  However, we are including the fermionic determinant in the
observables:  all the results presented here are obtained  with the
correct determinant (squared, for two flavors) weight.  From earlier
experience \cite{LaPa98b,FaLaWo98} we know that this is justifiable for
the presented statistics.  We perform our investigation on three sets
of 5000-10000 configurations at $\beta=2$, 4 and 6.

The so-called ``geometric definition'' of the topological charge is
\be\label{geocha}
Q_G=\frac{1}{2\pi}\,\sum_x\,{\rm Im\, ln}(U_{12}(x))\;;
\ee
we keep track of its value for all our configurations. The
configurations have been well separated by 
$3\,\tau_\textrm{\scriptsize int}$, the
autocorrelation length for $Q_G$.

For \DWI we need to determine $\kappa_c(\beta)$. We use PCAC
techniques for this purpose \cite{HiLaTe98} determining $\kappa_c$
for the unquenched 2-flavor case.

For each configuration we then build \DWI (at $\kappa_c(\beta)$), \DOV
and \DFP as discussed. Each lattice Dirac matrix is diagonalized
to obtain the complex eigenvalue spectrum.  This
is somewhat time-consuming due to the non-hermiticity. Furthermore the
inverse (the quark propagator) is determined.

In the 2-flavor Schwinger model one expects \cite{Co76} (for a recent
discussion cf. \cite{GaSe94}) one massive mode (called $\eta$ by
analogy) and a massless flavor-triplet (called $\pi$).  The
corresponding momentum-projected operators are
\begin{eqnarray}
\eta(p,t) &=& \sum_{x_1} \,e^{ipx_1} 
\left(\bar u(x_1,t)\,\sigma_1\,u(x_1,t)
+\bar d(x_1,t)\,\sigma_1 \,d(x_1,t)\right)\;,\\
\pi_3(p,t) & = &\sum_{x_1} \,e^{ipx_1} 
\left(\bar u(x_1,t)\,\sigma_1\,u(x_1,t)
-\bar d(x_1,t)\,\sigma_1 \,d(x_1,t)\right)\;.
\end{eqnarray}
Their correlation functions define by their exponential decay the
corresponding energy functions $E(p)$ and thereby the dispersion
relation. In the 1-flavor case only the massive mode is there.

We also study rotational symmetry via the correlation function
\begin{equation}\label{corfun}
P(x)=\langle\psibar(0)\,\sigma_3\,\psi(0)\,
\psibar(x)\,\sigma_3\,\psi(x)\rangle
\end{equation}
measured  for all 2-point separations.

As mentioned before, in order to avoid numerical problems with
inversions for the Dirac operator with (almost or exact) zero
eigenvalues we introduce a small regulator mass $\mu$. It turned out
that for $\mu={\cal O}(10^{-3})$ or smaller the result is practically
insensitive to this cut-off, the inversion algorithm still working
properly.

\section{Discussion of the Results}

\begin{figure}[ht]
\begin{center}
\epsfig{file=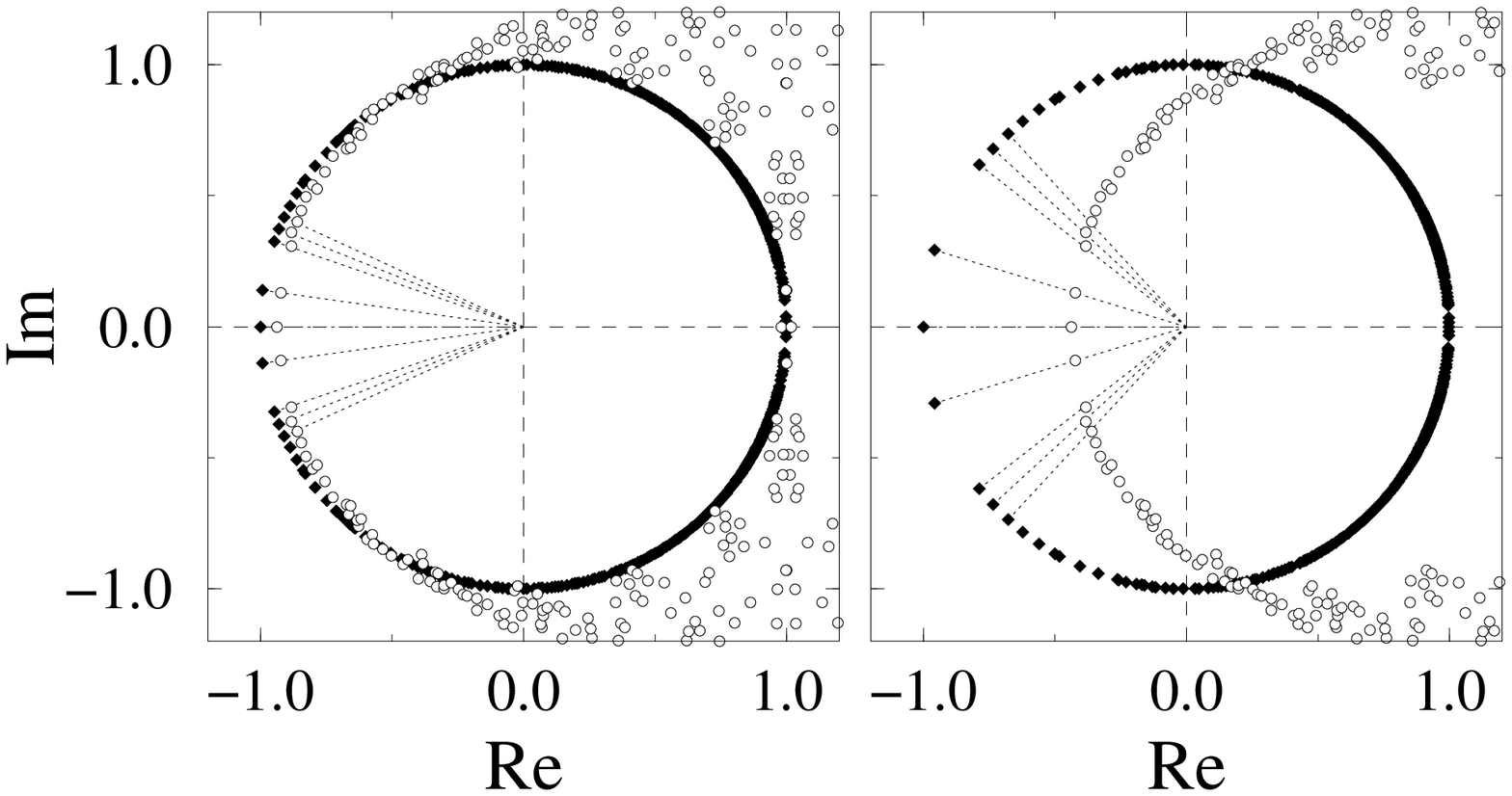,angle=0,width=12cm}
\end{center}
\vspace{-0.5cm}
\caption{\label{NeProj}
Eigenvalues of \DWI (open circles) and of the 
term $\gamma_5\, \epsilon(\gamma_5\,\DWI)$ (full diamonds) 
in \DOV, left: $\KNE=\frac{1}{2}$, right: $\KNE=\frac{1}{3}$.}
\end{figure}

From the definition (\ref{DOV}) of $\DOV$ one could naively expect that
its spectrum is obtained from the starting Wilson operator at a
proper value of $\KNE$ by simply projecting the eigenvalues onto the
circle on the complex plane $|\lambda-1|=1$.  This is not quite the
case, although it becomes more and more so for larger $\beta$
approaching the continuum limit. In particular the real modes are
projected either onto $\lambda=0$
or to $\lambda=2$ (see fig.\ref{NeProj}).  In
the case of a pair of real modes of opposite chirality, they split into
two conjugated complex eigenvalues on the circle.  Finally one is left
with zero modes of only one definite chirality -- in agreement with the
so-called vanishing theorem valid in D=2 \cite{VanTh} --
and an equal number of modes $\lambda=2$ with the opposite chirality.
As a consequence of the process of splitting of chirality pairs, the
number of zero modes in \DOV is smaller than those of real small
eigenvalues of \DWI.

In fig. \ref{NeProj} we compare -- for a fixed gauge configuration (at
$\beta=6$) -- the eigenvalues of $\DWI$ (at $\kappa=\frac{1}{2}$ and
$\frac{1}{3}$; n.b. this is still above $\kappa_c$) with those of
$\DOV-\unitmatrix=\gamma_5\,\epsilon(\gamma_5\,\DWI)$ resulting from the
projection (\ref{DOV}).  We find that the negative real eigenvalues of
\DWI are projected to $-1$ (corresponding to $\lambda=0$): their
number, counted according to the signs of their chirality, agrees with
the number of zero modes of \DOV.  In particular gauge configurations
for smaller $\beta$ have more eigenvalues on the real axis, with a
distribution density becoming broader when decreasing $\beta$.
Increasing (decreasing) $\KNE$ the spectrum of $\DWI$ would shift
towards left (right), and so more (less) zero modes might be obtained
as a result of the projection.  For small $\beta$ there is no clear
distinction for \DWI between the physical branch of the real spectrum
and the eigenvalues due to doubler modes.  This uncertainty is
discussed in the framework of the overlap method \cite{EdHeNa98} and
the Wilson- and Sheikholeslami-Wohlert action \cite{GaHi98}.

For \DOV, as for \DWI, not all zero modes can be equivalenced to the
geometrically defined topological charge of the gauge configuration.
Of course one can still {\em define} the topological charge as the
number of zero modes, or equivalently the number of $\lambda=2$ modes,
counting properly the chirality; in the case of the fixed point action
one obtains in this way the fixed point topological charge of the
lattice configuration \cite{HaLaNi98}.  This ambiguity vanishes towards
larger $\beta$, approaching the continuum limit.  Table
\ref{tabratioASIT} (upper part) shows the percentage $p(\beta)$ of
configurations where the number of zero modes counted according to the
sign of their chirality $(n_R-n_L)$ (cf. the discussion in \cite{He98})
agrees with the geometric topological charge.  The agreement of the
first two lines is trivially explained, since the real modes counted
for \DWI are those projected to $\lambda=0$
(or split into complex pairs in the case of a real pair of
opposite chirality) for \DOV.  This agreement stems from the choice of
$\KNE$  as ``cut-off'' for the counting of the real modes of \DWI.

\begin{table}
\begin{center}
\begin{tabular}{crrr}
\hline
Dirac op. & $p(\beta=2)$ & $p(\beta=4)$ & $p(\beta=6)$\\
\hline
\DWI &  74.22   &  99.70   & 100.00\\
\DOV &  74.22   &  99.70   & 100.00\\
\DFP &  96.58   & 100.00   & 100.00\\
\hline
\DWI &  62.56   &  97.36   &  99.78\\
\DOV & 100.00   & 100.00   & 100.00\\
\DFP &  91.00   &  99.84   &  99.96\\
\hline
\end{tabular}
\end{center}
\caption{\label{tabratioASIT} Upper part: percentage $p(\beta)$
of configurations where the Atiyah-Singer index theorem is fulfilled.
Lower part: percentage $p(\beta)$
of configurations where the vanishing theorem is fulfilled.}
\end{table}

In table \ref{tabratioASIT} (lower part) we confirm that for \DOV the
zero modes have just one definite chirality, whereas for the other
actions both chiralities contribute, displaying a violation of the
vanishing theorem (in the case of the \DFP we believe that this
violation is an effect of the truncation of less local couplings).  For
larger $\beta$ all three actions recover the vanishing theorem and the
ASIT: $(n_R-n_L)=Q_G$.

\begin{figure}[htp]
\begin{center}
\epsfig{file=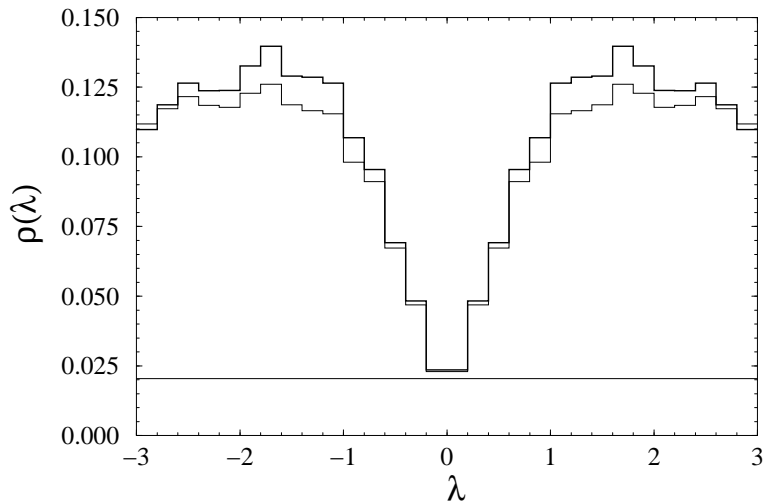,angle=-90,width=10cm}
\end{center}
\caption{\label{Density} The unquenched eigenvalue density distribution
projected from the circle onto the imaginary axis at $\beta=6$ for 
$\DFP$ (thick lines) and $\DOV$ 
with $\KNE=\frac{1}{2}$ (thin lines). The horizontal 
line denotes the continuum value at infinite volume.}
\end{figure}

The density distribution of eigenvalues on (for $\DOV$) or almost on
(for $\DFP$) the circle agrees with each other at small
eigenvalues, with improving agreement for increasing $\beta$.  For
$\KNE=\frac{1}{2}$ already at $\beta=6$ the densities are
indistinguishable within the statistical errors for
$\mbox{Im}(\lambda)<0.6$ (Fig. \ref{Density}, here we adopt the
definition in \cite{FaLaWo98} for the projection of the density 
distribution of eigenvalues to the imaginary axis). 
Choosing another $\KNE$ will produce another eigenvalue distribution. 
Comparing these with each other and with the value of
$\condensate$ \cite{BaCa80} one has to take care of the proper normalization
of the fermion fields, as discussed above in sec. \ref{sec:Neuberger}. 

A suitable lattice definition of $\condensate$ was suggested
in the framework of the overlap action \cite{Ne98d}, which can be
generalized for any fermions obeyed the GWC \cite{Ha98a} (see
\cite{FaLaWo98} for an application). Also this quantity shows
excellent agreement for $\DOV$ at $\KNE=\frac{1}{2}$ and $\DFP$.  We
obtain for the two actions (lattice units): 0.073(4) and 0.072(7)
respectively for $\beta=4$, 0.062(3) and 0.063(3) for $\beta=6$ (to be
compared to the continuum values 0.080 and 0.065 \cite{SaWi92}). Choosing
$\KNE=\frac{1}{3}$ with the appropriate normalization (with the factor
for free fermions) we find 0.058(1) at $\beta=6$.  The condensate has
been studied in the 1-flavor Schwinger model in the overlap formalism
already in \cite{NaNeVr95}, where also good agreement with the expected
values has been demonstrated.

\begin{figure}[htp]
\begin{center}
\epsfig{file=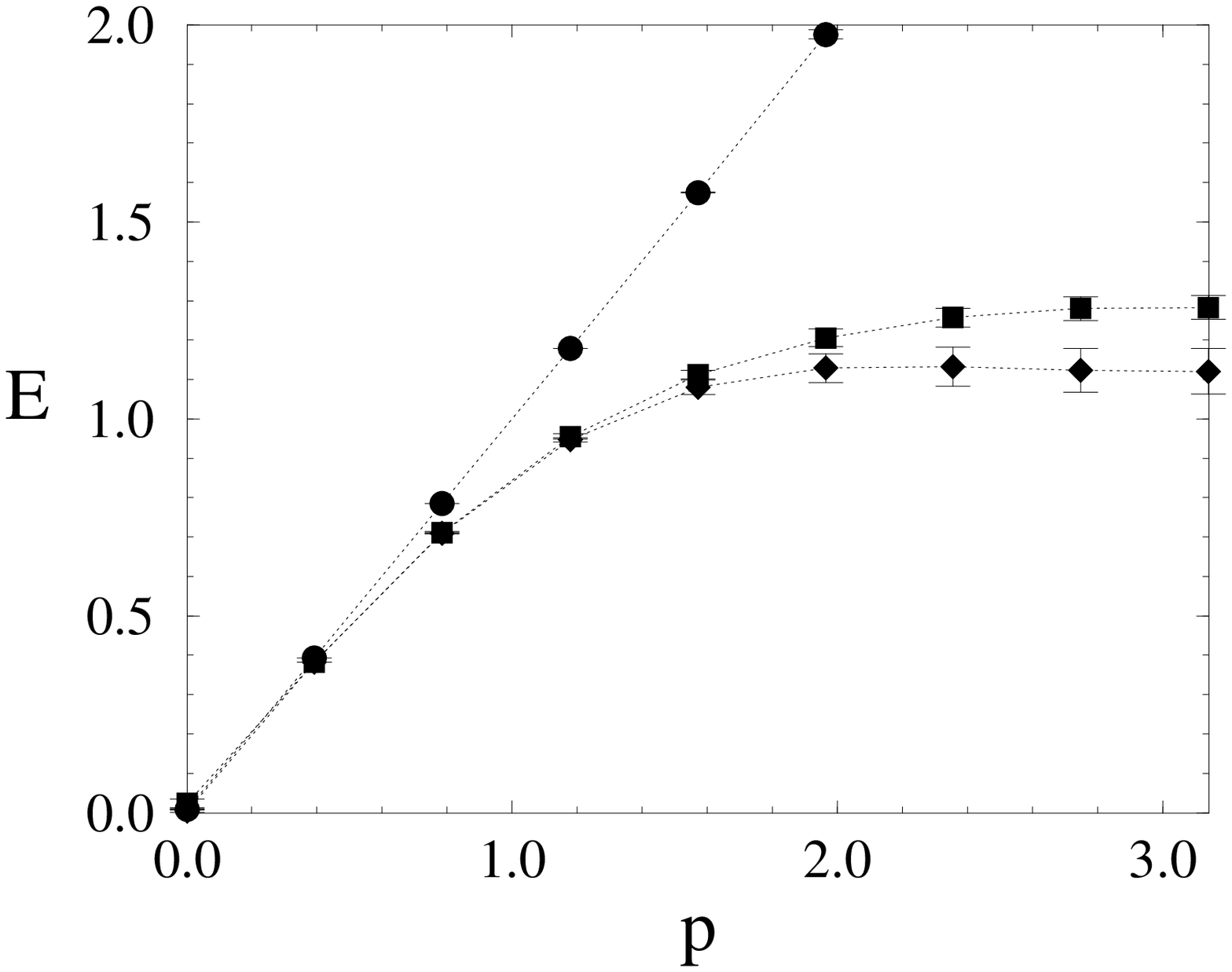,angle=0,width=7cm}\\
\epsfig{file=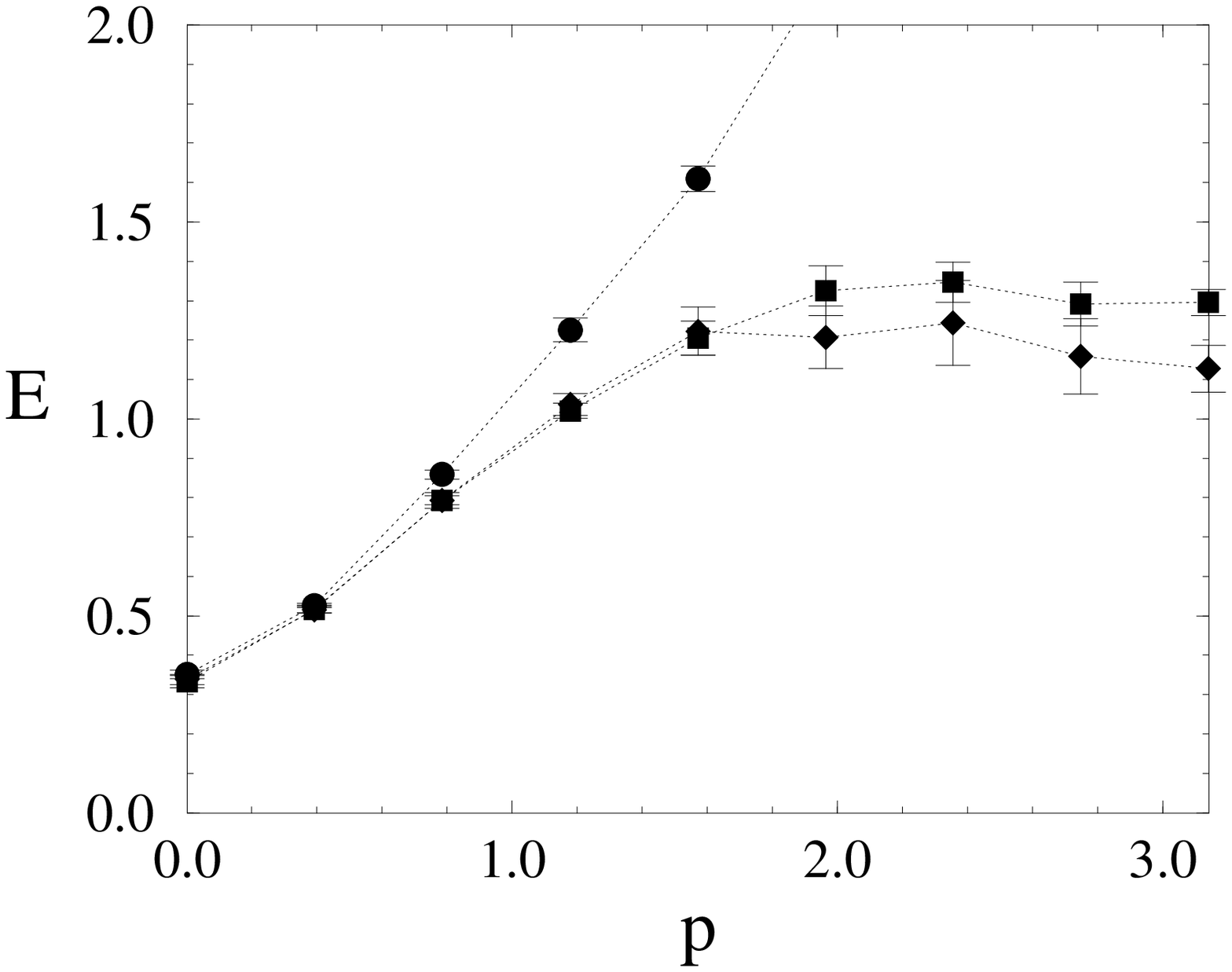,angle=0,width=7cm}
\end{center}
\caption{\label{DispRel}
The dispersion relation $E(p)$ for (upper figure) the $\pi$
and (lower figure) the $\eta$ propagator
for the three actions studied (lattice size $16^2$, $\beta=6$);
squares: Wilson action, diamonds: Neuberger action,
circles: fixed point action.}
\end{figure}

Discussing the mass spectrum, we find that for $\DOV$
($\KNE=\frac{1}{2}$) the vanishing of the $\pi$-mass is excellently
realized (see fig. \ref{DispRel}); this is an amazing improvement with
respect to \DWI, where the tuning to $\kappa_c$ creates non-irrelevant
technical problems.  The behavior of the energy dispersion for non-zero
momenta shows no improvement compared to \DWI, while for \DFP it is
almost linear as expected for a massless particle in the continuum:
\DFP eliminates the cut-off effects producing a continuum-like propagator
{\em for all momenta}. The massive state displays
qualitatively similar behavior. 

Decreasing $\KNE$ in the definition of \DOV from the suggested value
$\frac{1}{2}$ down to $\frac{1}{3}$ somewhat enlarged the
values of $E(p)$ slightly above those of \DWI, but still similar in the
overall shape. Changing further to even smaller $\KNE$ closer to
$\frac{1}{4}$ we observe, that the propagators do not reach asymptotic
behavior on the lattice size studied (even for
$\KNE > \kappa_c$) and no mass plateaus can be identified. We suspect,
that this indicates larger corrections to scaling.

We also studied  the correlation function (\ref{corfun}) for the  three
actions.  The rotational symmetry properties for \DOV are comparable to
those of \DWI. Action \DFP shows the best rotational invariance. In
summary, we find that the real-space correlations functions and the
dispersion relations are not noticeably improved by \DOV, as compared
to those for \DFP, which show a behavior substantially closer to the
continuum.

It is expected that \DOV is automatically ${\cal O}(a)$ corrected
\cite{KiNaNeNi98} and thus improves scaling for the on-shell quantities;
without at least introducing improvement of the current operators one
would not expect improvement for the propagators as exhibited by our
results.

We conclude that the identification of zero modes with the geometric
topological charge for \DOV agrees with that for \DWI, if the real
modes are counted according to their chirality. \DFP shows generally
better behavior. The vanishing theorem, however, is satisfied
automatically for \DOV, since only zero modes of one chirality result
from the projection.  The bound state masses and condensate come out
similarly.  The rotational invariance and the dispersion relations of
the current propagators are not significantly improved for \DOV as
compared to \DWI, but they are definitely better for \DFP.

\vspace*{0.5cm}
{\bf Acknowledgment:}

I.H. wishes to thank S. Chandrasekharan for a stimulating discussion
and for sharing some of his unpublished results.  We are grateful to
W.  Bietenholz and F.  Niedermayer for discussions.  Support by Fonds
zur F\"orderung der Wissenschaftlichen Forschung in \"Osterreich,
Project P11502-PHY is gratefully acknowledged.


\newpage
\begin{thebibliography}{10}

\bibitem{NiNi81}
H. Nielsen and M. Ninomiya,
\newblock Nucl. Phys. B 185 (1981) 20.

\bibitem{GiWi82}
P.~H. Ginsparg and K.~G. Wilson,
\newblock Phys. Rev. D 25 (1982) 2649.

\bibitem{Ha98c}
P. Hasenfratz,
\newblock Nucl. Phys. B (Proc. Suppl.) 63A-C (1997) 53.

\bibitem{NaNe939495}
R. Narayanan and H. Neuberger,
\newblock Phys. Lett. B 302 (1993) 62;
\newblock Phys. Rev. Lett. 71 (1993) 3251;
\newblock Nucl. Phys. B 412 (1994) 574;
\newblock ibid. B 443 (1995) 305.

\bibitem{HaLaNi98}
P. Hasenfratz, V. Laliena, and F. Niedermayer,
\newblock Phys. Lett. B 427 (1998) 125.

\bibitem{Lu98}
M. L{\"u}scher,
\newblock Phys. Lett. B 428 (1998) 342.

\bibitem{Ne98}
H. Neuberger,
\newblock Phys. Lett. B 417 (1998) 141.

\bibitem{Ne98a}
H. Neuberger,
\newblock Phys. Lett. B 427 (1998) 353.

\bibitem{LaPa98b}
C.~B. Lang and T.~K. Pany,
\newblock Nucl. Phys. B 513 (1998) 645.

\bibitem{FaLaWo98}
F. Farchioni, C.~B. Lang, and M. Wohlgenannt,
\newblock Phys. Lett. B 433 (1998) 377.

\bibitem{BaCa80}
T. Banks and A. Casher,
\newblock Nucl. Phys. 169 (1980) 103.

\bibitem{Ch98a}
S. Chandrasekharan,
\newblock hep-lat/9805015.

\bibitem{KiNaNe98}
Y. Kikukawa, R. Narayanan and H. Neuberger,
\newblock Phys. Rev. D 57 (1998) 1233.

\bibitem{NeEdHeNa98}
H. Neuberger,
\newblock hep-lat/9806025;
R.~G. Edwards, U.~M. Heller, and R. Narayanan,
\newblock hep-lat/9807017.

\bibitem{Ch98}
T.-W. Chiu,
\newblock hep-lat/9804016.

\bibitem{HiLaTe98}
I. Hip, C.~B. Lang, and R. Teppner,
\newblock Nucl. Phys. (Proc. Suppl.) 63 (1998) 682.

\bibitem{Co76}
S. Coleman,
\newblock Ann. Phys. 101 (1976) 239.

\bibitem{GaSe94}
C.~R. Gattringer and E. Seiler,
\newblock Ann. Phys. 233 (1994) 97.

\bibitem{VanTh}
J. Kiskis,
\newblock Phys. Rev. D 15 (1977) 2329;
N.~K. Nielsen and B. Schroer,
\newblock Nucl. Phys. B 127 (1977) 493;
M.~M. Ansourian,
\newblock Phys. Lett. 70B (1977) 301.

\bibitem{EdHeNa98}
R.~G. Edwards, U.~M. Heller, and R. Narayanan,
\newblock hep-lat/9802016.

\bibitem{GaHi98}
C. Gattringer and I. Hip,
\newblock hep-lat/9806032.

\bibitem{He98}
P. Hernandez,
\newblock hep-lat/9801035.

\bibitem{Ne98d}
H. Neuberger,
\newblock  Phys. Rev. D 57 (1998) 5417.

\bibitem{Ha98a}
P. Hasenfratz,
\newblock Nucl. Phys. B525 (1998) 401.

\bibitem{SaWi92}
I. Sachs and A. Wipf,
\newblock Helv. Phys. Acta 65 (1992) 653.

\bibitem{NaNeVr95}
R. Narayanan, H. Neuberger and P. Vranas,
\newblock Phys. Lett. B 353 (1995) 507.

\bibitem{KiNaNeNi98}
Y. Kikukawa, R. Narayanan and H. Neuberger,
\newblock Phys. Lett. B 399 (1997) 105;
see also:
F. Niedermayer,
\newblock plenary talk given at ``Lattice 98'', 
Boulder, July 1998.

\end{thebibliography}
\end{document}